# Annotation of Scientific Summaries for Information Retrieval

Fidelia Ibekwe-SanJuan[1], Silvia Fernandez[2], Eric SanJuan[2], Eric Charton[2]
[1] ELICO - University of Lyon3, France. {ibekwe@univ-lyon3.fr}
[2] LIA – University of Avignon
[2] {silvia.fernandez@univ-avignon.fr, eric.sanjuan@univ-avignon.fr, eric.charton@univ-avignon.fr}

**Abstract.**
We present a methodology combining surface NLP and Machine Learning techniques for ranking asbtracts and generating summaries based on annotated corpora. The corpora were annotated with meta-semantic tags indicating the category of information a sentence is bearing (objective, findings, newthing, hypothesis, conclusion, future work, related work). The annotated corpus is fed into an automatic summarizer for query-oriented abstract ranking and multi-abstract summarization. To adapt the summarizer to these two tasks, two novel weighting functions were devised in order to take into account the distribution of the tags in the corpus. Results, although still preliminary, are encouraging us to pursue this line of work and find better ways of building IR systems that can take into account semantic annotations in a corpus.

**Keywords.** Corpus annotation, discourse structure analysis, automatic summarization, document ranking, term weighting.

## 1. Introduction

The question of assisting information seekers in locating a specific category (facet) of information has rarely been addressed in the IR community due to the inherent difficulty of such a task. Indeed, efficiency and effectiveness have been the main guiding principles in building IR models and tools. Our aim here is to delve into the problem of how to assist a researcher or a specialist in rapidly accessing a specific category or class of information in scientific texts. For this, we need annotated corpora where relevant sentences are marked up with the type of information they are purportedly carrying. We identified eight categories of information in abstracts which can be useful in the framework of information-category driven IR: OBJECTIVE, RESULT, NEWTHING, HYPOTHESIS, FINDINGS, RELATED WORK, CONCLUSION, FUTUREWORK. These categories enable the user to identify what a paper is all about and what the contribution of the author is to his/her field. We adopted a surface linguistic analysis using lexico-syntactic patterns that are generic to a given language and rely on surface cues to perform sentence annotation from scientific abstracts. Once annotated, the corpus is fed into an automatic summarizer which takes into account the different semantic annotations for query-oriented document ranking and automatic summarization. The automatic summarizer used here is Enertex developed by LIA team at the University of Avignon (Fernández *et al*, 2007a). Enertex is based on neural networks (NN), inspired by statistical physics, to study fundamental problems in Natural Language Processing, like automatic summarization and topic segmentation.

In this paper, we will present some preliminary experiments on abstract ranking and automatic summarization using the semantic annotations resulting from our sentence categorization scheme.

The plan of this paper is as follows: section 2 recalls relevant related work; section 3 describes the sentence categorization method. Section 4 describes the query-oriented abstract ranking and automatic summarization experiments using the semantic annotations. Section 5 discusses difficulties inherent in this task as well as earlier unsuccessful experiments which we had attempted.

## 2. Related Work

Of a multi-disciplinary nature, our research draws from at least two distinct research communities: NLP and IR. Our survey will thus touch on relevant work from these two communities.

There is a large body of work in the NLP community on the structure of scientific discourse (Luhn 1958, Swales 1990, Paice 1993, Salager-Meyer 1990). Following a survey of earlier works, Teufel & Moens (2002) established that scientific writing can be seen as a problem-solving activity. Authors need to convince their colleagues of the validity of their research, hence they make use of rhetorical cues via some recurrent patterns (Swales 1990[1], Teufel & Moens 2002). According to Toefel & Moens (2002), meta-discourse patterns are found in

---

[1] « researchers like Swales (1990) have long claimed that there is a strong social aspect to science, because the success of a researcher is correlated with her ability to convince the field of the quality of her work and the validity of her arguments», cited in

almost every 15 words in scientific texts. It is thus feasible to present important information from sentences along these dimensions which are almost always present in any scientific writing: research goal, methods/solutions, results. Earlier studies also established that the experimental sciences respected more these rhetorical divisions in writing than the social sciences and more often than not, used cues to announce them. One of the goals of these studies has been and continues to be automatic summarization. Discourse structure analysis is a means of identifying the role of each sentence and thus of selecting important sentences to form an abstract. Teufel (1999), Teufel & Moens (2002), and then Orasan (2001) have pursued this line of research. Patterns revealing the rhetorical divisions are frequent in full texts but are also found in abstracts. For instance, within the division « *Motivation/objective/aim* », one could find the sentence containing the lexico-syntactic cue « *In this paper, we show that...* ». Teufel & Moens (2002) showed that authors took great pains in abstracts to indicate intellectual attribution (references to earlier own work or that of other authors). Since abstracts contain only the essential points of a paper, it is to be hoped that only important sentences are there and that therefore their classification is an easier task than classifying sentences from full texts. However, abstracts will not carry all the patterns announcing the different rhetorical divisions. While categories like objective, methods and results will almost always be present, others like "*new things, hypothesis, related_work, future_work*" may be missing.

Research on automatic summarization *per se* has become very dynamic of late. Sparked off by Luhn in the late 50's (Luhn 1958) who developed a system of sentence extraction, automatic summarization is the process that transforms a source text into a target, smaller text in which relevant information is condensed. Different techniques have been explored for this task. They can roughly be split into two broad families: those relying primarily on NLP and those relying primarily on statistical / machine learning models. Quite often, a combination of techniques from the two families is necessary to produce satisfactory summaries. The dominant approach to remains automatic summarization by sentence selection rather than by real abstraction, using statistical models to rank sentences according their relevance (Mayburi Mani, 1999). Some post-processing using NLP techniques is usually needed to smoothen the most glaring coherence problems.

The works of Teufel & Moens (2002) and Orasan 2001 can be classified in the NLP-oriented approach. Teufel & Moens (2002) developed a system called Argumentative Zoner for detecting the rhetoric function of sentences according to a detailed classification of rhetoric patterns in English. They trained a Naïve Bayes classifier to categorize sentences in 80 full text scientific articles from the computational linguistics field. This classifier attained an accuracy of (73%) in classifying sentences according to the different categories of information they announced. Basing on the work of Teufel & Moens (2002), Genoves *et al*. (2007) developed the AZEA authoring tool (Argumentative Zoning for English Abstracts) to identify the discourse structure of scientific abstracts. These authors also used machine learning techniques (decision trees, Naïves Bayes, rule learning algorithm and SVM) to categorize sentences from 74 abstracts from the pharmacology domain. The SVM classifier attained the highest degree of accuracy (80.3%) on well structured abstracts. This performance dropped to 74.8% when abstracts written by learners (students) were considered.

The majority of automatic summarization systems are based on statistical and/or machine learning models. Among the criteria and techniques explored, we can cite textual position (Edmundson 1969; Brandow *et al*. 1995; Lin and Hovy 1997), Bayesian models (Kupiec *et al*., 1995), SVM (Mani and Bloedorn, 1998; Kupiec *et al.*, 1995), maximum marginal relevance (Goldstein *et al.*, 1999). These studies also take into account structural information from the document such as benchmark words and structural indicators (Edmundson, 1969; Paice 1990), a combination of information retrieval and text generation to find patterns or lexical strings in the text (Barzilay and Elhadad, 1997; Stairmand, 1996). Automatic summarization systems can also be viewed alongside the number of documents summarized at a time: single or multi-documents. Lately, the focus has been on multi-document summarization. However, at least three challenges face multi-document summarization: redundancy removal, novelty detection and detection of contradictory information. The first two problems are of course related. For the elimination of redundancy, current studies rely on temporal cues in documents. A general method for addressing novelty detection lies in extracting the temporal labels such as dates, past periods or temporal expressions (Mani and Wilson 2000) or in building an automatic chronology from the literature (Swan and Allan, 2000). Another technique that uses the well-known position of $\chi 2$ (Manning and Schütze, 1999) is used to extract unusual words and phrases from the documents. A study comparing redundancy removal techniques (Newman *et al*., 2004) showed that a similarity measure like the cosine measure between sentences attained a similar performance to other more complex methods such as Latent Semantic Indexing (LSI) (Deerwester *et al.,* 1990). Research on automatic summarization has come a long way since its beginning. Despite the residual problem of lack of coherence and cohesion, the summaries proposed by automatic systems are an approximation of the human summary.

---

Teufel & Moens (2002: p. 413).

Our approach to sentence categorization and query-oriented abstract ranking and summarization combines the two major techniques from the NLP and machine learning communities. We first perform sentence categorization by building on earlier works on the discourse structure of scientific texts (Teufel & Moens 2002). Like in Teufel[2] (1999), we adopt a domain-independent level of linguistic analysis. The major goal of these authors was to build summaries in such a way that the new contribution of the source article can be situated with regard to earlier works. This is in line with the recent task of "novelty detection" in multi-document summarization which was added in the last "Document Understanding Conferences" (DUC[3]) challenge. After annotating the corpus with the different categories of information each sentence contains, we perform query-oriented abstract ranking and automatic summarization. This part is done with the Enertex system, an automatic summarizer based on the neural networks approach inspired by the statistical physics of magnetic systems. Enertex is based on the concept of «textual energy». The principal idea behind Enertex is that a document can be viewed as a set of interactive units (the words) where each unit is affected by the field created by the others. The algorithm models documents as neural network whose interaction or "textual energy" is studied. Because of the nature of the links that the measure of energy induced, it connects to both sentences with common words and sentences that are in the same vicinity without sharing necessarily the same vocabulary. Textual energy has been used as document similarity measure in NLP applications. What makes this system more interesting is its ability to handle quite different tasks. In principle, the textual energy can be used to score sentences in a document and separate those that are relevant from those that are not. This led immediately to a strategy of single-document summarization by extracting phrases (Fernández *et al.*, 2007a). On the other hand, using a query as an external field in interaction with a multi-document corpus, we have broadened the scope of this idea to develop an algorithm for query-guided summaries (Fernández *et al.*, 2007b). So we calculated the degree of relevance (the textual energy) of the corpus sentences to the query. Query-guided summaries have been evaluated in the context of DUC's tasks. Enertex system compares very favorably to the other participating systems because, in essence, textual energy is expressed as a simple product matrix. Another less obvious application, is to use the information of this energy (seen as a spectrum of the sentence) and compare it with others. This allows the detection of thematic boundaries in a document. For this comparison we used the test match between Kendall. Enertex attained performances equivalent to state of the art (Fernández *et al.*, 2007a). Here, we have adapted it to the task of query-oriented abstract ranking taking into account semantic annotations present in the corpus and in the queries.

## 3. Lexico-syntactic patterns acquisition for sentence categorization

### 3.1 Corpora

To determine the type of information carried by each sentence, we need to identify and characterise the patterns that introduce that particular information type. We have selected eight categories of information which a user can seek for in scientific discourse in the framework of novelty detection: objective, results, newthings, findings, hypothesis, future work, related work, conclusions. To acquire patterns reflecting the eight categories of information we want to mark up, we analyzed corpora from three different disciplines. The 1st corpus was made up of 50 abstracts on Quantitative biology from the Open Archives Initiative (OAI[4]) containing the word '*gene'*. We manually read and analyzed the first 50 abstracts in order to formulate our initial set of patterns, seen as the seed patterns. The seed patterns were then automatically projected onto two other corpora using Unitex linguistic toolbox, in order to test their portability and to acquire new patterns. Thus, pattern acquisition was done incrementally. The second corpus consisted of 1000 titles and abstracts from 16 Information Retrieval journals downloaded from the PASCAL[5] database. The third corpus from the field of Astronomy, was made up of 1293 titles and abstracts from the ISI[6] Web of Science (WoS) database, containing the the term "Sloan Digital Sky Survey" (SDSS[7]). We describe below in more details how the initial set of patterns acquired manually from the first corpus were implemented and projected onto the two remaining corpora.

---

[2] He spoke of steering "*clear of distinctions that are too domain specific*", adding that it was necessary to take into account "*robustness requirements of our approach, we cannot go indefinitely deep: the commonalities we are looking for must still be traceable on the surface*" (ibid, p.83).
[3] duc.nist.gov/guidelines/2007.html
[4] http://fr.arxiv.org/archive/q-bio
[5] http://www.inist.fr.
[6] Institute for Scientific Information
[7] http://www.sdss.org/

## 3.2 Implementation of the patterns as finite state automata

Lexico-syntactic patterns announcing a specific information type are not fixed expressions. They are subject to variations. These variations can occur at different linguistic levels: morphological (gender, number, spelling, inflection), syntactic (active/passive voice, nominal compounding *vs* verbal phrase), lexical (derived form of the same lemma) and semantic (use of synonymous words). The exact surface form of all these variations cannot be known in advance. Hence, categorizing sentences based on these surface patterns requires that we take into account places where variations can occur so as to ensure that they can be applied to new corpora with a certain degree of success. From our manual study of the 50 abstracts in Quantitative biology, we wrote contextual rules in the form of regular expressions implemented as finite state automata in the Unitex[8] system. These automata were then projected on the two test corpora to identify the different categories of sentences. Verbs are searched for in their infinitive form, nouns in their noun masculine gender.
Figure 1 below shows the finite state automaton that recognize OBJECTIVE sentences.

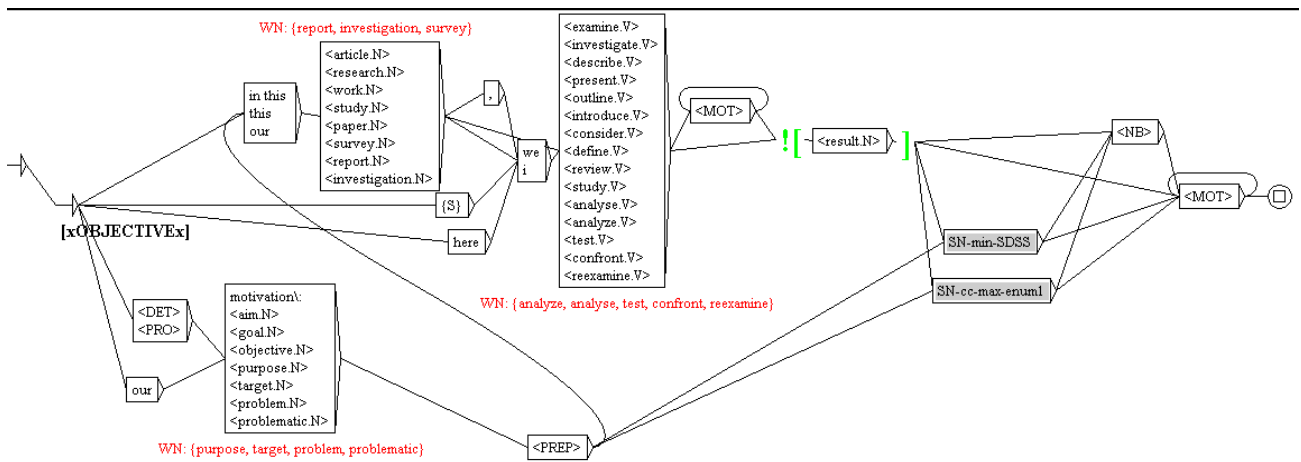

Figure 1. Finite state transducer that categorizes sentences as «OBJECTIVE».

The recognition of the patterns require the combined use of POS[9] and lexical information and syntactic-level information (recognition of noun phrases in the context of a lexical pattern). In Figure 1, the pattern for identifying objective sentences contains a path which searches for a sentence with a determiner (<DET>) or a pronoun (<PRO>), followed by words like "*goal, objective, purpose, aim*,..." then by a preposition (<PREP>), and by a noun phrase (SN-cc-max-enum1). The grey boxes call other finite state grammar embedded in the current one. For instance, «SN-cc-max-enum1» is a local grammar that identifies complex NPs (NPs with embedded simpler NPs). This grammar in turn, embeds another simpler NP grammar. The expressive power of such local grammars can be quite high as simpler grammars are embedded into more complex ones to achieve a considerable level of complexity. Each category of information is represented by a single automaton with multiple paths. Note however that some lexico-syntactic patterns are ambiguous and can introduce two different categories of information. For instance, there is not always a clear boundary between patterns announcing the objective of a paper and its results. "*In this paper we show that...*" could announce either "*objective"* or "*results*". Genoves *et al*. (2007) observed that the classifiers they trained could not distinguish properly between "methodology" and "results" patterns.
To ensure the completeness of our lexico-syntactic patterns and hence their portability on other domains, we expanded the lexical lists in the patterns with words in the semantic equivalence classes from an external lexical database, in this case WordNet[10]. However, WordNet being a general vocabulary semantic resource, has every conceivable sense for a given word, some of which were not appropriate for scientific writing. For instance, the verb "*show*" has among its synsets the following "*render* (sense of picture), *read*, *register*, *evince*" which are senses rarely encountered in scientific writing. A second unwelcome phenomenon in expanding word lists with WordNet is that if word $w_0$ has as synonyms word $w_1$, there is no guarantee that the synonyms of word $w_1$ will be synonyms of word $w_0$. In other words, synonymy is neither always symmetric nor transitive. For instance, among the synonyms of *"obtain"*, is the word *"receive"*, the latter has synonyms like *"welcome, meet, pick up"* which are clearly not synonyms *"obtain"* in the sense used in scientific articles. Of the total of 9506 sentences in the SDSS corpus, 1 882 (19%) unique sentences were tagged by our restricted patterns and 1 959 (20%) sentences by the expanded patterns with WordNet, thus the coverage by adding lexical entries (synonyms) from an external resource was not significantly increased.

---

[8]   www-igm.univ-mlv.fr/~unitex/
[9]   Part-Of-Speech
[10]  http://wordnet.princeton.edu/perl/webwn

### 3.3 Corpus annotation

Once the patterns have been built and tested, the second stage is to mark-up sentences in the corpus with the category of information they announce. This is done by using the transducer option in Unitex. Transducers are variants of the grammars that modify the text by performing a re-writing operation such as "insert, delete, copy". The information carried by each pattern is inserted at the beginning of the sentence containing the pattern. Figure 2 shows an example of the output by the transducers of each local grammar. The tags [OBJECTIVE, RESULT, HYPOTHESIS] were inserted by our finite state grammars.

---

{S}ISI:{S}000240201200022.
{S}Potential sources of contamination to weak lensing measurements: constraints from N-body simulations.
[OBJECTIVE] {S}We investigate the expected correlation between the weak gravitational shear of distant galaxies and the orientation of foreground galaxies, through the use of numerical simulations.{S} [HYPOTHESIS] This shear-ellipticity correlation can mimic a cosmological weak lensing signal, and is potentially the limiting physical systematic effect for cosmology with future high-precision weak lensing surveys.{S} We find that, if uncorrected, the shear-ellipticity correlation could contribute up to 10 per cent of the weak lensing signal on scales up to 20 arcmin, for lensing surveys with a median depth z(m) = 1.{S} The most massive foreground galaxies are expected to cause the largest correlations, a result also seen in the Sloan Digital Sky Survey.{S} [RESULT] We find that the redshift dependence of the effect is proportional to the lensing efficiency of the foreground, and this offers prospects for removal to high precision, although with some model dependence.{S} The contamination is characterized by a weakly negative B mode, which can be used as a diagnostic of systematic errors.{S} We also provide more accurate predictions for a second potential source of error, the intrinsic alignment of nearby galaxies.{S} This source of contamination is less important, however, as it can be easily removed with distance information.

---

Figure 2. Example of an annotated abstract.

Next, we evaluated the accuracy of our sentence tagging grammars by manually verifying the output of the different automata on the SDSS corpus. Each sentence was read in order to ascertain if it really belonged to that particular category of information. The table below gives figures on the accuracy of the each automaton in annotating sentences from a given category. The 2[nd] column is the total number of sentences tagged by an automaton. The 3[rd] column gives the ratio of correctly tagged sentences over all tagged sentences (precision). The 4[th] column is the proportion of errors amongst sentences tagged. Recall could not be measured because we could not read the entire corpus to exhaustively identify all the sentences belonging to a specific category that were not tagged. In the future, we plan to measure recall on a sample of the corpus. The automaton for hypothesis sentences embeds the one for "finding" because the two categories of information are often announced by similar patterns. This explains why we have seven patterns in the table instead of the eight announced previously.

| Pattern | Occ. | Prec. | Errors |
|---|---|---|---|
| RESULT | 500 | 100% | 0 |
| CONCLUSION | 206 | 193 (94%) | 13 (6%) |
| FUTURE_WORK | 198 | 194 (98%) | 4 (2%) |
| NEWTHING | 505 | 485 (96%) | 20 (4%) |
| OBJECTIVE | 513 | 513 (100%) | 0 |
| RELATED_WORK | 31 | 30 (97%) | 1 (3%) |
| HYPOTHESIS | 487 | 479 (98.4%) | 8 (1.6%) |

Table 1. Accuracy measure of the automata for tagging sentences on the SDSS corpus.

As we can see, our patterns achieved a high level of accuracy in tagging sentences with the correct type of information (> 94%). The majority of the errors observed in the conclusion sentences came from the fact that the word "conclusion" or "conclude" which are triggers for tagging a sentence as such were present in the sentence but the actual conclusion came in the following sentences (see appendix A for examples). A possible way of correcting this would be to extend the conclusion class to the "*n*" sentences following the one containing that word. The majority of the errors observed in the hypothesis-findings categories come from recommendations using the trigger word "*should*" or from future work using the word "*shall*". Positive and negative examples of sentences tagged for each category of information can be found in the appendix. For two categories of patterns (objective, result), we could find no error in the tagged sentences. This might be due to the highly technical nature of the SDSS corpus. We might observe more errors in a less technical corpus.

### 3.4. Automatic pattern generation

A limit of the rule-based approach which we have adopted here for pattern acquisition and sentence tagging is that it is impossible to capture all the potential patterns especially in unseen texts. Previous studies used machine learning techniques to address this issue. Teufel & Moens 2002, then later Genoves *et al*., 2007 trained classifiers on manually hand-crafted patterns. However, the authors trained the classifiers on the same corpus as the initial one used to build the patterns in order to evaluate their accuracy. They did not actually use them to learn new patterns.

As a first step to new patterns acquisition, we applied a rule generator in order to systematically generate all the possible lexical combinations of words in similar contexts in the patterns in the eight categories. Here is a detailed description of the algorithm.

To find the generated patterns, we use a substitution class. Consider the sample class of substitution called « *demonstrate* » from the result category (left box in figure 3 below). Our algorithm will first locate all the occurrences of each term of the « *demonstrate* » class in the corpus, and associates them with *n words* before or *n words* after. For example with n=2, and only word after, ee could find in the corpus the patterns in the middle box. For the purpose of this presentation, let us call those extended patterns *P1*. In a second step, we will substitute in P1, all the terms of the *demonstrate* class. This will give us a new list of candidate patterns. If we achieved this on P1 by substitution of the « *demonstrate* » class, we could have a proposition list, called P1' (the rightmost box).

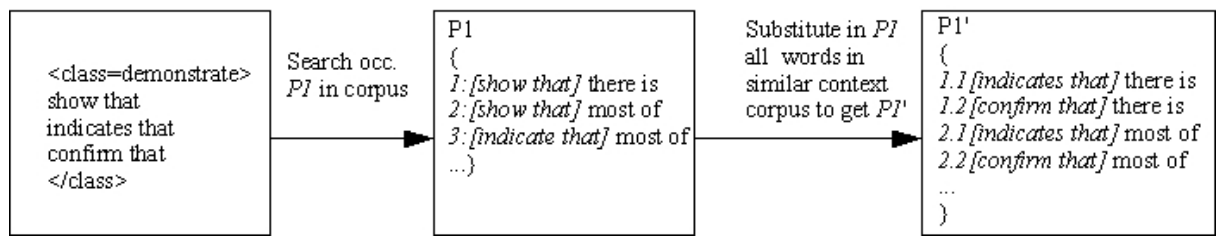

Figure 3. Flowchart of the pattern generator.

We see in P1', that 1.1 and 1.2 are substitutions of the *demonstrate* class terms, applied on the sequence 1 from P1. The rule generator checks for the presence of these new patterns in the corpus and records their occurrences. This ensures that our sentence classification program will not miss any sentence carrying a specific information type owing to lexical variations by substitution of one synonymous word with another. Owing to lack of time, we were not able to apply the generated patterns on the SDSS corpus in order to evaluate the accuracy and number of annotated sentences. This will be the object of future research.

## 4. Query-oriented abstract ranking

In this section, we explore how the semantic tags inserted in the abstracts (cf. §3.2 - 3.3) can be used for query-oriented abstract ranking and multi-abstract summarization. Enertex was selected as an appropriate tool because of its ability in capturing non-direct relations between queries and abstracts. We implemented a new combination of weighting functions in the Enertex system specifically for these tasks. One of the additional advantages of this system is its ability to handle quite different tasks of text selection and ranking with minor changes. First, we give a general description of the system.

### 4.1. Text representation in the Enertex system

The system builds large matrices *M* of word occurrence in a collection of small texts and computes a similarity between texts based on $(M.M^t)^2$. This is the matrix representation of the energy in the magnetic Ising model (Fernández et al, 2007a). Documents are pre-processed following conventional methods. First, functional, stop words and numbers are filtered out. Then normalization and lemmatization of words are carried out to reduce the dimensionality. A *bag-of-word* representation of the texts is performed, yielding a matrix $M=[f(w,s)]_{w \in W, s \in S}$ of weighted frequencies consisting of a set $S$ of $P$ sentences (lines) and a vocabulary $W$ of i = $1, \cdots, N$ terms (columns), where $f$ is a weighting function on pairs of words and sentences. We use some elementary notions of the graph theory to describe Enertex approach.

Let us consider the sentences as sets $S$ of words. These sets constitute the vertices of the graph. We draw an edge between two vertices $s,t$ every time they share at least a word in common. We obtain the graph $I(S)$ from intersection of the sentences. We weigh these pairs $\{s, t\}$ which we call edges using the weighting binary function $f$ on pairs of words and sentences :

$$e(s,t) = \sum_{w \in s \cap t} (f(w,s) \times f(w,t))$$  [Equation 1]

For automatic summary, $f(u,s) = 1$ if $u$ is in $s$, and 0 otherwise. In this special case, $e(s,t)$ is the number $|s \cap t|$ of words that share the two connected vertices. Finally, we add to each vertex $s$ an edge of reflexivity $\{s\}$ valued by the sum of weights $f(u,s)$ of words $u$ in sentence $s$.

This weighted intersection graph is isomorphic with the adjacency graph $G(M \times M^T)$ of the square matrix $M \times M^T$. In fact, $G(M \times M^T)$ contains $P$ vertices. There is an edge between two vertices $s,t$ if and only if $[M \times M^T]_{s,t} > 0$. The matrix of Textual Energy $E$ is $(M \times M^T)^2$. This matrix is computed using its adjacent graph whose vertices are the same as those of the intersection graph $I(S)$ and:
- there is an edge between two vertices each time that there is a path of length 2 in the intersection graph;
- the value of an edge: a) loop on a vertex $s$ is the sum of the squares of the values of adjacent edges at the vertex, and b) between two distinct adjacent vertices $r$ and $t$ is the sum of the products of the values of the edges on any way with length 2 between both vertices. These ways can include loops.

From this representation, it can be seen that the matrix of Textual Energy connects at the same time sentences sharing common words because it includes the intersection graph as well as sentences in the same neighborhood but not necessarily sharing the same vocabulary. Thus, two sentences $s, t$ not sharing any word in common but for which there is at least one third phrase $r$ will be connected all the same. The strength of this link depends in the first place on the number of sentences in its common neighborhood, and on the vocabulary appearing in a common context. This constitutes the main distinction with other usual similarity measures like cosine or mutual information measures that are based on direct co-occurrences of terms. Therefore, Textual Energy is comparable to Latent Semantic Indexing (LSI) without requiring the expensive computation of the Singular Value matrix decomposition. The advantage of such a model is that they allow to directly use query terms that appear only once in the corpus but that are closely related to some central topics. Since Textual Energy is based on a simple graph model, it is more adaptable to different applications. In text summarization, it was tested on several corpus including DUC 2006 and DUC 2007. Its performance was measured with ROUGE metrics and it showed similar performances to other state of the art systems (Fernandez *et al.*, 2007b ).

## 4.2. Ranking abstracts with semantic annotations

We now describe how Enertex was adapted to the task of ranking abstracts. All the experiments here were performed on the SDSS corpus. First, each abstract is considered as a unique *bag-of-words* (a sentence). In other to take into account the frequency of words in abstracts and to favor low frequency words that best characterize an abstract, we used the following weighting function $f$ on pairs of words and sentences based on the so-called "equivalence index" which is the product of the conditional probabilities P(s/w) and P(w/s). Only values over a threshold of the form $10^{-n}$ where $n$ depends on the corpus size are considered. Thus we set:

$$f(w,s) = \log(trunc((\frac{f_{w,s}^2}{f_{w,.} \times f_{.,s}} > 10^{-n}) \times 10^n))$$  [Equation 2]

where $f_{w,s}$ is the absolute frequency of word $w$ in $s$, $f_{w,.}$ is the frequency of $w$ in the corpus and $f_{.,s}$ is the number of words in sentence $s$. To optimize the ranking algorithm, we truncated float numbers to work only on integers and we cut too big values using the log. We tested the common versions of TF.IDF measures but due to double matrix product involved in the calculation of the Textual Energy matrix, the results showed an exaggerated effect of any weighting on the $S$ matrix favoring tacitly the extreme cases (long or short phrases; frequent or infrequent terms). Sentences are then ranked based on their weighted degree in the adjacent graph of $(M \times M^T)^2$: the score of a sentence $s$ is set to the sum of $E_{s,t}$ for any sentence $t$.

The system selects the most representative abstracts and displays them in chronological (by publication date). If two abstracts have the same score, only the first one by chronological order is displayed.

When ranking abstracts in response to a query, the query *q* is considered as an abstract itself. The corpus of abstracts is ranked according to the $E_{s,q}$ value. If the query contains a very general word then the ranking is similar to the one obtained without query.

By way of example, let us consider the query: "*Randall-Sundrum*". This term is the name of a space geometry model which occurred only once in the SDSS corpus. Using the above defined weighting function, Enertex ranked the abstract containing "*Randall- Sundrum*" and those dealing with geometry models. Enertex found the relationship between the named entity in the query and the geometry models based on the context in which it found the query term. Examples of relevant terms in this context are *geometry, spatially flat, dimension, inflation, expansion, brane, braneworld, DGP model*. This is similar to a query expansion procedure in which terms from the top ranked abstracts are used to expand the query term. The difference here is that Enertex selects the top ranked abstracts to expand the query based on the adjacent graph of the Energy matrix. Figure 4 shows one of these abstracts ranked on 7[th] position. Relevant terms are underlined.

> Two new one-parameter tracking behavior dark energy representations omega=omega(0)/(1+z) and omega=omega(0)e(z/(1+z))/(1+z) are used to probe the geometry of the Universe and the property of dark energy. The combined [RESULT] type Ia supernova, Sloan Digital Sky Survey, and Wilkinson Microwave Anisotropy Probe data indicate that the Universe is almost spatially flat and that dark energy contributes about 72% of the matter content of the present universe. The observational data also tell us that omega(0)similar to-1. It is argued that [FINDING] the current observational data can hardly distinguish different dark energy models to the zeroth order. The transition redshift when the expansion of the Universe changed from deceleration phase to acceleration phase is around z(T)similar to 0.6 by using our one-parameter dark energy models.
>
> Figure 4. One of the top ranked abstracts for the query "*Randall-Sundrum*".

Figure 4. Astract ranked 7[th] for the query "*Randall-Sundrum*".

If now we want Enertex to take into account the semantic annotation inserted into the abstracts following the connections in the same adjacent graphs. A difficulty we have to deal with here is that by definition, if the summaries follow the hypothesis of well-formedness, each semantic category tag will tend to be uniformly distributed across the corpus and will therefore have a high occurrence. Thus, when considered as words, the tags are simply ignored by the weighting function. To overcome this handicap, we multiplied our weighting function by a *g* factor that measures by how much the frequency of a word is greater than the expected one. Due to the corpus size, we could not apply complex statistical tests and most of the calculus had to be done on integers. Finally, we tried the following function:

$$g(w,s)=\log\left(trunc\left(\frac{(f_{w,s}-\overline{f_{w,.}}>0)^2}{\sum_{t\in S}(f_{w,t}-\overline{f_{w,.}})^2}\times f_{w,.}\right)\right) \quad \text{[Equation 3]}$$

This function compares the frequency of a word or a tag to the average frequency of this item in abstracts. Only items above the average are considered as index of abstracts. Therefore this function allows us to also consider some frequent tags as abstract index .We combine the two functions *f* in Equation 1 and *g* in Equation 2 by taking their product: (f(u)+1).(g(u)+1) if at least one of the two terms in not null (f(u)+f(g)>0) to obtain a ranking that both considers specialized terms in query and general tags.

For example, we added semantic tags to the previous query "*Randall-Sundrum NEWTHING FINDING*". The results showed that this combination effectively allows the system to rank abstracts according to these two principles. Abstracts containing the query terms are still ranked first but those containing an unusual number of tags in the query are favored. Figure 5 shows some sentences of abstract ranked on 19[th] position. It contains at the same time terms like "*dimensional*" related with "*Randall-Sundrum*" and "*FINDING*". Relevant terms are underlined. In the previous case, without any tag in the query, the same abstract had been ignored.

> Overall, the galaxy spectral energy distribution in the entire ultraviolet to [FINDING] near-infrared range can be described as a single-parameter family with an accuracy of 0.1 mag, or better. This nearly one-dimensional distribution of galaxies in the multidimensional space of measured parameters strongly supports the [CONCLUSION] conclusion of Yip et al., based on a principal component analysis, that [FINDING] SDSS galaxy spectra can be described by a small number of eigenspectra.

Figure 5. Some sentences from an abstract ranked 19[th] for the query "*Randall-Sundrum NEWTHING FINDING*".

Table 2 shows the differences between these two queries according to the content of some terms related to *"Randall-Sundrum"* and *NEWTHING FINDING* tags. It presents the percentage of ranked abstract where they appear. We observe that the percentage of related terms is almost the same and the percentage of tags used in the query increases significantly.

| Query | Some terms related with *Randall-Sundrum*: geometry, spatially flat, dimension, inflation, expansion, brane, braneworld, DGP model | Tags: *NEWTHING, FINDING* |
|---|---|---|
| ***Randall-Sundrum*** | 37% | 57% |
| ***Randall-Sundrum NEWTHING FINDING*** | 30% | 88% |

Table 2. Percentage of ranked abstract where terms related with *Randall-Sundrum* and *NEWTHING FINDING* tags appear.

We emphasize that with this ranking function, we do not need to specify which words are tags and which are terms. The ranking function naturally distinguishes them based on their frequency. This functionality should guarantee a high stability of the resulting rankings even when the annotation of texts is incomplete. Since we used a unifying model, the system should automatically learn from the context in which existing annotations appear to process other texts that should have been annotated in a similar way. We plan to evaluate this stability property on partially enriched texts from heterogeneous sources. The final system that we target, will rely on a:

1. tagging using the finite state automata introduces in section 3.2 to partially tag texts. This tagging appears to have a high precision but since we cannot evaluate its recall, we shall consider it as partial.
2. learning process based on the automatice pattern generation introduced in section 3.4 that shall detect text features of succeeding text related to tags.
3. text analysis based on Textual Energy that can relate abstracts to queries made up of any list of terms including those appearing once in the text collection and tags relying on a precise but partial tagging.

## 4.3. Query-oriented multi-abstract summarization using semantic annotations

Here, we focus on query-oriented multi-abstract summarization. In this context, summaries are a selection of documents' abstracts (instead of sentences) displayed by chronological (publication) order. To better evaluate the ability of the system to capture non-direct relations between queries and abstracts, and to determine the impact of the semantic tags in the queries, we present two types of experiments. The first one involves one-word queries consisting of abbreviations or of astronomy concepts. The aim is to see if the system was capable of producing summaries that contained a definition of the abbreviation and some related information. The experiment will consist of queries with and without tags to study their impact in summary content. The second experiment consists of phrase-tagged queries describing a phenomenon or problem related to astronomy. In this case, the aim is to observe if the summary gives information that helps to explain the problem raised in the query. The different tags will be added too to give a predominant intention to the summary.

*4.3.1 One-word queries*
Consider a set of four one-word queries consisting of abbreviations or of astronomy concepts (Table 3), the aim was to see if the system was capable of producing summaries that contained a definition of the abbreviation and related information. Given a compression rate $r$, the system selects the top most ranked abstracts such that the total number of their words over the total corpus size is less than $r$. We fixed the compression rate of the resulting summary to <5 % of the corpus size in terms of total number of words. The SDSS corpus is made up of 258 775 words. This induces that summaries produced by the system can contain different numbers of abstracts depending on their size in words. If two abstracts have almost the same score, they are considered redundant and only the most recent is selected. If all abstracts have a null score because no one could be related to the query, the summary will be empty. Therefore, the length of the summary also depends on the number of abstracts with a non null score.

We present now a preliminary evaluation of our approach. First we check that Textual Energy is sufficient to relate queries to abstracts. Until now, Textual Energy was used to rank sentences in which a term rarely appeared twice, meanwhile here we consider abstracts. To evaluate its performance, we consider query terms with very low frequencies but that are acronyms involving relevant topics of the corpus. Based on established definitions of these acronyms, the evaluation consisted in counting the number of relevant terms in these definitions that appear in the abstracts selected by the system. This done, we will enrich the query with tags and see if the document ranking is modified or if it fails because of the high frequency of tags.

Table 3 shows the four one-word queries, their occurrence in the corpus, the size of the generated summary in number of words. For ease of comprehension, we added the definitions of the query terms and indicated the websites from where they were taken.

| Id | Query | Corpus occ. | Nb. words summary | Definition of query term |
|---|---|---|---|---|
| b1 | ACDM | 5 | 0 | ΛCDM or Lambda-CDM is an abbreviation for Lambda-Cold Dark Matter. |
| b2 | AGB | 2 | 9690 | Asymptotic Giant Branch. http://www.eso.org/projects/vlti/science/node8.html |
| b3 | AMIGA | 2 | 9679 | Analysis of the interstellar Medium of Isolated GAlaxies http://amiga.iaa.es:8080/p/1-homepage.htm |
| b4 | LBG | 3 | 9692 | Lyman break galaxies http://www.astro.ku.dk/~jfynbo/LBG.html |

Table 3. Examples of one-word abbreviation queries tested and their definitions.

From the results, it appeared that for small values of $n$ (<4) in equation 2, only terms of very low frequency (<5) are retained. Since we carried out the experiments with n=4, query b1 (*ACDM*) did not produce any abstract.

Analysing the contents of the abstracts selected to build the summary for query b2 "*AGB*", we note:
- that the summary contains 48 abstracts;
- the presence of the scientific term used in the query (*AGB*);
- the presence of scientific terms present in the query term's presentation on the website such as *Asymptotic Giant Branch* (2 occurrences in the summary), *Life* (2), *Core* (5), *Non-LTE* (4), *Convection atmosphere* (3), *stratosphere* (3), *chemical evolution* (1).

Enertex was again able to find the relationship between the named entity in the query and the related concepts based on the context in which it found the query term.

Similarly, query b3 "*AMIGA*" produced a summary of 31 abstracts. Comparing this summary with the persentation of the term on a website (address in table 3), we found the following terms in common: *amiga* (2), e*nvironment* (29), *interaction* (3), *correlation* (15), *environmental density* (1), *isolated galaxy* (7), *denser environments* (15), *wavelength* (3), *Catalog of Isolated Galaxies* (1).

Finally, query b4 (*LBG*) produced 34 abstracts that share with the website presentation the following terms: *LBG* (7), *Ultraviolet* (3), *Red* (in the rigth context: 1), *Lyman* (8), *Rest-frame* (1), *Ly-α* (2)

Another important observation is that terms specific to queries b2 and b3 like *AGB, lte* and a*miga* are not present in the summary of b3. Meanwhile more general terms relavant to b2 and b3 like "*isolated galaxy"* and "*denser environements*" occured also in the summary of b4 but with a much lower frequency (1 and 2 respectively).The summaries were obtained using the product (f($w,s$)+1).(g($w,s$)+1) of two formulas in Equations 2 and 3. However, the second factor (g($w,s$)+1) did not influence the ranking, this factor being equal to 1 for all query terms $w$ and all abstracts $s$.

Seeking to determine the impact of semantic tags in the query, we added the tags announcing hypothesis, findings and objectives to the queries. We observed that all produced summaries are non null, even for query b1. This is due to the effect of the second factor $g$. To illustrate this, let us take a closer look at the results produced for query b2 "*AGB*". Similar observations can be made for the other queries. Relevant terms in the presentation of b2 mentioned in table 3 are less frequent but still present. In table 4, we can see that the importance of scientific terms related to "*AGB*" that were present in the presentation mentioned in table 3 has declined but are still present. On the other hand, the total number of tags for "hypothesis, finding and objective" are higher in the selected abstracts.

| Term occurrence in the summary | query without tags | query with tags |
|---|---|---|
| agb | 2 | 1 |
| life | 2 | 1 |
| core | 5 | 3 |
| non-LTE | 4 | 0 |
| convection atmosphere, stratosphere | 3 | 0 |
| chemical evolution | 1 | 1 |
| hypothesis | 4 | 15 |
| finding | 8 | 19 |
| objective | 19 | 15 |

Table 4. Frequency of relevant terms and of tags in summary produced for query b2 "*AGB*".

*4.3.2 Phrase-tagged queries*

The aim here is to evaluate to what extent the summary gives information to explain the problem raised in the query and if the use of tags orients the predominant intention in the generation of the summary. An example is the query **"***NEWTHING spectral classification of quasar"*. Table 5 shows the percentage of ranked abstracts where terms related to the query and semantic annotations are present. Relevant terms appeared in all abstracts forming the summary. The tag **"***NEWTHING"* was more present than others tags.

| **Terms related** *(luminosity, quasar, redshift, quasar spectra, spectrum, optical~, Balmer, eigen~, Fe II emission, Baldwin)* | NEWTHING | RESULT | CONCLUD | HYPOTHES | OBJECTI | FINDING |
|---|---|---|---|---|---|---|
| 100% | 72% | 60% | 16% | 20% | 48% | 24% |

Table 5. Percentage of ranked abstract where terms query related (first column) and tags appear for the query *"NEWTHING spectral classification of quasar"*.

Figure 6 shows some of the sentences taken from an abstract ranked 1$^{st}$ and 14$^{th}$ respectively. Terms relevant to the query are underlined.

> [NEWTHING] We found that more infrared *luminous* galaxies tend to have a smaller local galaxy density, being consistent with the picture where luminous IRGs are created by the merger-interaction of galaxies that happens more often in lower-density regions.
>
> We find strong correlations between the [NEWTHING] detection fraction at other wavelengths and optical properties such as flux, colours and emission-line strengths.

Figure 6. Some sentences of ranked abstracts for the query **"***NEWTHING spectral classification of quasar"*. Relevant terms and tags are underlined.

In another query, the phrase *"existence of the Gunn-Peterson"* was entered in combination with different semantic tags. We did an evaluation of system's effectiveness by measuring again the presence of the relevant terms in the summary. We have identified as relevant terms related with *"existence of the Gunn-Peterson"*: *neutral hydrogen, intergalactic medium, IGM, detection+existence, quasar spectra, Lyman+Alpha, z=5.99,6.28, reionization.* The results are shown in Table 6. We observe that relevant terms are always very present in the summary and the use of a tag in the query favors his presence in the final condensed.

| Id | Query | Terms related with *"existence of the Gunn-Peterson"* | **HYPOTHESIS** | **FINDING** | **CONCLUD** | **RESULT** |
|---|---|---|---|---|---|---|
| p2 | *HYPOTHESIS existence of the Gunn-Peterson* | 93% | **43%** | 17% | 17% | 35% |
| p3 | *FINDING existence of the Gunn-Peterson* | 84% | 24% | **48%** | 24% | 56% |
| p4 | *CONCLUSION existence of the Gunn-Peterson* | 89% | 18% | 29% | **33%** | 37% |
| p5 | *RESULT existence of the Gunn-Peterson* | 89% | 22% | 33% | 22% | **44%** |

Table 6. Query *"existence of the Gunn-Peterson"* in combination with different tags.

# 5. Discussion

Regarding the sentence classification task, we observed that the same patterns can announce different information categories or that two different patterns can be present in the same sentence, thus leading to multiple tags. In the following sentence, the future_work tag is triggered by the word « *future* » while the hypothesis tag is triggered by the presence of "*can*":

« We assess the accuracy with which [xHYPOTHESISx] [xFUTURE_WORKx] future galaxy surveys can measure *cosmological parameters.* »

Teufel & Moens (2002) already observed the same phenomenon on a different corpus which was from computational linguistics. In this case, the sentence will belong to the two classes as there is no clear way of determining which category should take precedence.

Concerning the sentence ranking and automatic summarization tasks, we first tried to generate query-oriented multi-abstracts summaries by sentence selection. The results were not satisfactory because the extracted sentences lacked sufficient context to be coherent. Moreover, the resulting ranking of sentences was similar to a random ranking. The use of semantic tags in queries did not change this outcome as if the sentences did not contain enough informatiion to be related to queries. We next tried to use the weighting function in Equation 3 to generate summaries from DUC 2006 corpus. The generated summaries had lower quality scores for ROUGE measures than those obtained without using this weighting function. This shows that ranking abstracts is a different task from ranking full-text documents. We then tried ranking sentences from this corpus but encountered the same problem as previously.

It was then we had the idea of working at the level of abstracts. At this level, query terms can be related to similar terms appearing in the same abstracts but in different sentences. In a sentence, a word relevant to the query typically appears once whereas this is not the case in abstracts. Because the common versions of TF.IDF did not produced the expected effects, it was thus necessary to define special weighting functions that captured low frequency terms in the corpus but which are more frequent in a smaller set of abstracts. This gave rise to the function proposed in Equation 2. This formula could not take into account semantic tags that are both frequent and uniformly distributed in all abstracts. Indeed, any well written scientific abstract would tend to contain at least one category of patterns from the major rhetorical divisions (objective, method, results, conclusion). However, some abstracts can contain an unusual number of these patterns and this information could be relevant for document ranking. Equation 3 is meant to capture such unusually high frequency of rhetorical patterns in abstracts.

Finally we found out that working at the abstract level, it was possible (as described in section 4) to define weighting functions that can take into account both rare terms and frequent semantic tags considered as supplementary words in the text. This opens an avenue for research where standard IR engines could, with minor changes, be applied on annotated corpus.

Enertex was initially designed for automatic summarization by sentence extraction and text segmentation. It attained performances equivalent to state of the art summarizers and segmentation systems. Here, we have adapted it to the task of query-oriented abstract ranking taking into account semantic annotations present in the corpus and in the query. We have to pursue these experiments in oder to determine the best way of focusing the generated summaries or rankings on a specific information type. Also, we have to set up a more rigorous evaluation framework using corpora with benchmarked results such as the DUC collections. However, what makes this system most interesting is its ability to handle quite different tasks of text selection and ranking with minor changes.

This work had a double purpose. First it shows an easy way to tag peer-reviewed abstracts according to the information carried by each sentence. Second it shows how tags can be used in a text analysis process with the view to perform automatic summarization. Text analysis tasks are part of information retrieval, they rely on reduced document collections extracted from large databases using standard Information Retrieval methods but requiring a higher level of text understanding. The methods we developed in this work constitute a novel and integrated approach for addressing advanced information retrieval tasks.
.

**Appendix.**

Examples of positive and negative sentences tagged by the automata for sentence classification

| Pattern | Pos_example | Neg_example |
| --- | --- | --- |
| Results | Model comparisons indicate that the age of the young population of these galaxies does not vary with radius.<br>We find that the slope of composite LFs becomes flatter toward a redder color band.<br>We find that the spectral classification of quasars is redshift and luminosity dependent; | |
| Conclusions | Hence, we claim the possible universality of the color of the galaxies on the red sequence.<br>Therefore, the existence of the Gunn-Peterson trough by itself does not indicate that the quasar is observed prior to the reionization epoch.<br>We therefore conclude that the point source is likely to be a fifth lensed image of the source quasar. | With this large sample, we have reached the following conclusions.<br>Our analysis leads to the following conclusions:<br>Our findings are as follows.<br>One method is to search for gaps in the Gunn-Peterson absorption troughs of luminous sources. |
| Future_work | Further host galaxy observations will be needed to refine the significance of this result.<br>We emphasize the need for further observations of SNe in the | I will review some of the latest developments on cosmological reionization and suggest, in a somewhat more personal way, that the universe may |
| | rest-frame UV to fully characterize, refine, and improve this method of SN type identification. Future work needed to extend this selection algorithm to larger redshifts, fainter magnitudes, and resolved sources is discussed. | be reionized twice in order to paint... No other planned survey will provide so much photometric information on so many stars. The full SDSS data set will include greater than or similar to 1000 SDSS/RASS clusters. |
| Newthing | In this paper we report the discovery of a new X-shaped radio galaxy with a partially obscured quasar nucleus. We present evidence for eight new clumps of blue horizontal branch stars discovered in a catalogue of these stars compiled from the Sloan Digital Sky Survey by Sirko et al. and published in 2004. The OLS-lens survey: the discovery of five new galaxy-galaxy strong lenses from the SDSS. | However, only more extensive optical photometry and a detection of its spin or spin-orbit beat frequency can confirm this classification. Detection of quasar clustering anisotropy would confirm the cosmological spacetime curvature that is a fundamental prediction of general relativity. Here we present the New York University Value-Added Galaxy Catalog (NYU-VAGC), a catalog of local galaxies ( mostly below z approximate to 0.3) based on a set of publicly released surveys matched to... |
| Objective | This paper describes spectra of quasar candidates acquired during the commissioning phase of the Low-Resolution Spectrograph of the Hobby-Eberly Telescope. We present results from 1 month, 3 year, and 10 year simulations of such surveys. We investigate the luminosity dependence of quasar clustering, inspired by numerical simulations of galaxy mergers that incorporate black hole growth. | |
| Related_work | In contrast to past findings, we find that not all M7 - M8 stars are active. Our results are in excellent agreement with recent determinations of these relations by Mandelbaum et al. using galaxy-galaxy weak lensing measurements from the SDSS. Unlike previous work, however, we are able to detect structures in the lens associated with cluster galaxies. | This distribution has been found to have fractal dimension, D, approximately equal to 2.1, in contrast to a homogeneous distribution in which the dimension should approach the value 3 as the scale is increased. |
| Hypothesis | Knowing that all three methods can have significant biases, a comparison can help to establish their (relative) reliability. A combination of all three effects may better explain the lack of Lyalpha absorption reduction. A larger sample of QSO pairs may be used to diagnose the environment, anisotropy, and lifetime distribution of QSOs. We estimate that the SRN background should be detected (at 1sigma) at Super-K in a total of about 9 years ( including the existing 4 years) of data. | Redshifts may have been assigned to some QSOs due to misidentification of observed lines, and unusual spectra should be particularly investigated in this respect. This estimate is based on small-sample statistics and should be treated with appropriate caution. The revision should be taken into account in any future analysis of the source number density of UHECRs based on the ORS. |